\documentclass[journal]{IEEEtran}
\usepackage[utf8]{inputenc}

\ifCLASSINFOpdf
   \usepackage[pdftex]{graphicx}
\else
   \usepackage[dvips]{graphicx}
\fi
\usepackage{amsmath}
\usepackage{amsfonts}
\usepackage{optidef}

\usepackage{algorithmic}

\usepackage{lineno}
\usepackage{color}

\IEEEpubid{\begin{minipage}{\textwidth}\ \\[60pt]
        \centering\small{	© 2021 IEEE. Personal use of this material is permitted. Permission from IEEE must be
obtained for all other uses, in any current or future media, including
reprinting/republishing this material for advertising or promotional purposes, creating new
collective works, for resale or redistribution to servers or lists, or reuse of any copyrighted
component of this work in other works.}
\end{minipage}} 

\begin{document}
%
\title{Frequency Selection for Platoon Communications in Secondary Spectrum Using Radio Environment Maps}
%
%
%

\author{Marcin Hoffmann,~\IEEEmembership{Student~Member,~IEEE,}
        Pawel Kryszkiewicz,~\IEEEmembership{Senior~Member,~IEEE,}
        and~Adrian~Kliks,~\IEEEmembership{Senior~Member,~IEEE}
\thanks{The work has been realized within the project no. 2018/29/B/ST7/01241 funded by the National Science Centre in Poland.

The authors are with the Institute
of Radiocommunications, Poznań University of Technology, Polanka 3, 61-131 Poznań, Poland e-mail: marcin.ro.hoffmann@doctorate.put.poznan.pl.}
}

%
%

\markboth{IEEE Transactions on Intelligent Transportation Systems (T-ITS)}%
{Hoffmann \MakeLowercase{\textit{et al.}}: Frequency Selection for Platoon Communications in Secondary Spectrum Using Radio Environment Maps}
%



\maketitle

\begin{abstract}
Platoon-based driving is an idea that vehicles follow each other at a close distance, in order to increase road throughput and fuel savings. This requires reliable wireless communications to adjust the speeds of vehicles. Although there is a dedicated frequency band for vehicle-to-vehicle (V2V) communications, studies have shown that it is too congested to provide reliable transmission for the platoons. Additional spectrum resources, i.e., secondary spectrum channels, can be utilized when these are not occupied by other users. Characteristics of interference in these channels are usually location-dependent and can be stored in the so-called Radio Environment Maps (REMs). This paper aims to design REM, in order to support the selection of secondary spectrum channel for intra-platoon communications.
We propose to assess the channel's quality in terms of outage probability computed, with the use of estimated interference distributions stored in REM. A frequency selection algorithm that minimizes the number of channel switches along the planned platoon route is proposed. Additionally, the REM creation procedure is shown that reduces the number of database entries using (Density-Based Spatial Clustering of Applications with Noise) DBSCAN algorithm. The proposals are tested using real IQ samples captured on a real road. Application of the DBSCAN clustering to the constructed REM provided 7\% reduction in its size. Utilization of the proposed channel selection algorithm resulted in a 35 times reduction of channel switches concerning channel assignment performed independently in every location.

\end{abstract}


%
\IEEEpeerreviewmaketitle

\section{Introduction}
%
%
%
%

\IEEEPARstart{T}{he} number of vehicles on the roads consistently grew, through recent years, causing increasing traffic congestion. This implies not only time delays but also energy wastes and pollution. An obvious solution to improve road throughput is to deploy additional infrastructure, e.g., by building new roads or extending the number of lanes of existing ones.
However, this solution is first time-consuming, secondly expensive, and finally, it is not always possible to expand vehicular infrastructure e.g., to build additional lane under urban conditions. On the other hand, there is an idea to deal with the mentioned issues by changing the driving pattern into the so-called \emph{platoon}-based~\cite{HALL2005405}. \emph{Platoon}-based driving pattern assumes that a group of vehicles will follow each other, while maintaining short inter-vehicles distance, e.g., few meters~\cite{jia2016}. The first vehicle in the platoon is the platoon-leader. It is typically responsible for the management, and coordination of platoon behavior, e.g., adjusting speed, or inter-vehicle distance. However, also distributed solutions are under consideration~\cite{zheng2019}. There are several benefits from the deployment of the \emph{Platoon}-based driving. First, short inter-vehicle spaces improve the road capacity that is also related to the reduction of traffic congestion~\cite{vanArem2006}. Secondly, vehicles following the platoon leader save fuel, implying the reduction of carbon footprint~\cite{chan2012}. The shorter the inter-vehicle distance, the higher the expected fuel savings. However, very short inter-vehicle distance requires precise and error-free coordination of vehicles in order to prevent collisions. 

This is one of the key challenges in \emph{platoon}-based driving, i.e., to ensure safety for vehicles. Each car within the platoon must be able to adapt its speed to the platoon leader, including sudden breakings, while maintaining short inter-vehicles distance. Studies have shown that for this purpose it is more beneficial to rely on the short-range wireless communications sending, e.g., a message about breaking, than on direct measurements from distance sensors~\cite{xu2014}. The short-range wireless communication between vehicles within the platoon can be realized using, e.g, Dedicated Short-Range 
communications (DSRC). The DSRC physical and medium-access layers are described in IEEE 802.11p, and Wireless Access in Vehicular Environment (WAVE) standards~\cite{Abdelgader2014ThePL}. WAVE operates in the dedicated frequency band: 5.850-5.925 GHz, and 5.855-5.925 GHz in the USA, and Europe, respectively. This is one of the Unlicensed National Information Infrastructure (U-NII) bands. However, studies have shown that with a growing number of vehicles utilizing DSRC this amount of spectrum will not be sufficient to provide low enough latency, and high reliability~\cite{REDDYG2018720, bohm2013}. Moreover, similar issues could be identified while considering platoon communications based on the cellular technologies e.g. Cellular Vehicle-to-Everything (C-V2X)~\cite{wang2019}.
The reason for this is the limited capacity of the wireless channel. While the intra-platoon spectrum access could be designed in an orthogonal, collision-free manner, the orthogonality between transmissions of various platoons and vehicles cannot be guaranteed. Even if different platoons use the same waveform, for which orthogonality can be obtained, it will not be achieved as a result of the lack of coordination, e.g., in general, different platoons are not time-synchronized and can start transmissions in the same time instance.
This lack of orthogonality results in inter-platoon interference, which reduces wireless channel capacity and causes communication delays. These delays may further lead to, e.g., instability in platoon formation, or reduced cyber security~\cite{souza2020, harfouch2018,basiri2019}. This would further imply larger inter-vehicle distances, that will significantly reduce the \emph{platoon}-based driving benefits, e.g. fuel saves.  

From this perspective, it seems reasonable to find an alternative frequency band that can be utilized for inter-platoon communications. Because vehicles within the platoon already formulate a small network, the whole communication can be synchronously offloaded to the less occupied frequency band. Such a procedure cannot be easily applied for general V2V communications, because it would require special synchronization mechanisms. 
One solution is to offload the platoon communications into the unlicensed band, e.g., 2.4~GHz, where no protection of other users' transmission is necessary. Although this approach seems not to be appropriate in the urban areas, where wireless access points density is large, it looks promising to be utilized under highway conditions~\cite{GHANDOUR20132408}. In~\cite{kim2011} a so-called Cognitive Anypath Vehicular Protocol (CoRoute) was proposed, in order to utilize instantaneous channel state information, and neighboring vehicles positions for opportunistic vehicular communication over $2.4~GHz$ industrial, scientific, and medical (ISM) band. 
Authors of \cite{wang2018} focus on the coexistence of V2X and Vehicular ad-Hoc network (VANET) systems in the unlicensed band, proposing a spectrum sensing scheme, vehicle interference models, and resource allocation algorithm. 
However, these works do not consider location-dependent interference characteristics.

Another idea is to look for the additional spectrum resources in the millimeter-waves band~\cite{choi2016, giordani2017}. However, in such high frequencies radio channel, the achievable range is very limited, which decreases the connection reliability. Reliability, that is crucial for platoon stability. 

On the other hand, although almost all of the frequency bands below 3 GHz are assigned to a variety of wireless systems, they are not fully utilized in practice~\cite{kliks2013}. Therefore, there is a possibility to exploit additional spectrum resources in the moments when these are not occupied by Primary Users (PUs), i.e., wireless systems that are licensed to transmit in a given band. This scheme is known in the literature as Dynamic Spectrum Access (DSA)~\cite{zhao2007}. DSA can further benefit from the information about the surrounding radio environment e.g., interference. It has been shown that this kind of information is in most cases related to the location, which makes an opportunity to create intelligent geolocation databases, known as the Radio Environment Maps (REMs)~\cite{yilmaz2013}. One example of REMs application is a storage of information about the unoccupied television channels: so-called TV White Spaces (TVWS)~\cite{kliks2017,yin2012,beek2012}. It is applicable because the terrestrial TV signal is stable over time, and location, i.e., location of the terrestrial TV transmitters along with their transmission power and channels assignment remains constant over a long-time period. 

Most importantly, the DSA, e.g., TVWS, is a well-suited solution for intra-platoon communications.  
The main reason is the low potential of interference caused by a platoon to other services. This is caused by relatively small transmission range, rapid changes of platoon position and other wireless services typically distanced from the road sites. 
Various REM architectures and implementation challenges for the purpose of DSA in platoon communications using TVWS had been described in~\cite{sybis2018}. In~\cite{sroka2020} authors proposed an algorithm to exploit TVWS data stored in REM, to apply DSA in V2X communications. 
While the above-mentioned REM applications assume relatively stable spectrum conditions and fixed PU types, REM can be used to represent more sophisticated cases as well, e.g., for scenarios with time-varying radio conditions and heterogeneous PUs. However, in this case, the PUs-originating interference has to be monitored and modeled to allow for the required reliability of intra-platoon communications.

The aim of this work is to propose a REM-based method of selecting alternative frequencies, where intra-platoon communications can be transferred. The main focus is put on the assessment of the radio channels quality under complex interference patterns. 
We expect that offloading intra-platoon communications to the high-quality channel, e.g., of low outage probability, will improve the general performance of control algorithms e.g. Cooperative Adaptive Cruise Control (CACC). While the platoon control mechanisms are out of the scope of this paper, the proposed, advanced models of interference can be applied to improve platoon stability and safety modeling done in~\cite{basiri2019, harfouch2018}.
 The final goal of this paper is an optimal frequency assignment to the platoon. The REM contains location-dependent information about interference in a given frequency band. Various radio bands can be characterized by various interference patterns, varying over time. We propose to characterize interference at a given channel and location as a random variable, and REM to store proper interference distribution parameters. This solution allows for higher accuracy than typically used single-number interference characterization, e.g., by its mean power~\cite{katagiri2018,sybis2018, sroka2020}. As the interference distribution at a given location will be estimated using a limited number of sensing reports, its accuracy can be insufficient. On the other hand, the interference properties in the neighboring locations should be closely correlated. As a result, some interference distributions in adjacent locations should be merged in order to increase accuracy. At the same time, REM size can be reduced. For this purpose, we propose to modify a Density-Based Spatial Clustering of Applications with Noise (DBSCAN) algorithm. The proposed modification aims to incorporate both geographical neighborhood and interference distribution similarities in the clustering procedure. Moreover, we propose a channel quality assessment method utilizing the interference distributions stored in REM. This channel quality assessment method is necessary to assign reliable enough channel for platoon communications. We propose to assign a wireless channel to the platoon by minimizing the number of frequency switches along the planned route while providing sufficiently low channel capacity outage. An optimal solution is proposed utilizing the well-known Dijkstra algorithm~\cite{Mehlhorn2008}.
The test field for validation of our proposal is the 2.4~GHz band. This part of radio frequencies is occupied by a variety of systems, e.g. WiFi, Bluetooth, ZigBee, providing various and non-regular interference patterns. Through the analysis of real data captured in the 2.4~GHz band during the measurement campaign, we will show that these interference patterns can be accurately represented using Gaussian Mixture Model (GMM).

 The main contribution of the paper are:
\begin{itemize}
    \item We propose an outage probability as a metric for channel quality assessment. The advantage of this metric over, e.g., latency is universality. Moreover, it can be obtained relatively easily, and irrespective of the transmission scheme, and medium access algorithms used. State-of-the-art solutions, e.g.,~\cite{zhang2019}, rely on creating location-dependent Signal-to-Interference-and-Noise-Ratio (SINR) maps. This approach makes those maps valid only for devices of equal transmission power and does not reflect the influence of narrowband interference on the signal. In our solution, we separate interference distribution from other parameters, e.g., pathloss and transmission power. This enables flexibility of providing frequency band to the platoon on the basis of platoon-specific requested transmission parameters.  
    \item We propose to train Gaussian Mixture Models (GMMs) in order to model interference distribution in secondary spectrum channels. State-of-the-art Radio Environment Maps usually are designed so as they contain pairs of location-mean interference power~\cite{katagiri2018, sybis2018, sroka2020}. This approach is sufficient while modeling interference from relatively stable sources, e.g., terrestrial television stations. However, such a model cannot be applied to scenarios where many interference sources of various transmission schemes operate simultaneously. Under such conditions more advanced models are necessary, i.e., the proposed (GMMs).
    \item We propose to utilize the DBSCAN clustering algorithm in order to reduce REM size. Among other state-of-the-art clustering algorithms, DBSCAN has an ability to take into the account both close geographical distance between REM entries and similarities between interference distributions. Prior work related to the topic of REMs did not consider such an approach~\cite{katagiri2018, sybis2018, sroka2020}. The procedure of DBSCAN clustering can be thought of as a process reverse to kriging~\cite{roger2020}. Instead of interpolating between REM points, we propose to create clusters (areas) where interference follows the same distribution. 
    \item We propose to use the Dijkstra algorithm to reduce the number of channel switches along the planned platoon route. We represent the problem of minimization of channel switches as a graph, where nodes refer to locations and edges represent available radio channels. A similar approach was proposed for the planning of UAV path~\cite{zhang2019}. However, in our case due to the nature of the platoon’s route, the computational complexity is much lower. Another state-of-the-art channel assignment algorithms~\cite{chen2011, kuldeep2018} aim at lowering the number of channel switches in the vehicular communication scenario but without forecasting the whole platoon route. As a result, they will not guarantee the performance of our, Dijkstra-based algorithm.    
\end{itemize}

 The rest of the paper is organized as follows: Sec.~\ref{sec:rem_architecture} describes REM deployment, channel assessment method, minimization of channel switches along the platoon route using Dijkstra algorithm, and adaptation of DBSCAN algorithm to enable REM size reduction. The field measurement campaign setup, together with recorded data distribution analysis is provided in Sec.~\ref{sec:measurement_campaign}. Evaluation of REM algorithms using the measurement data is described in Sec.~\ref{sec:simulation}. Conclusions are formulated in Sec.~\ref{sec:conclusions}.

\section{Radio Channel Quality Assessment utilizing Radio Environment Maps} \label{sec:rem_architecture}
%

\begin{figure}[!t]
\centering
\includegraphics[width=3.5in]{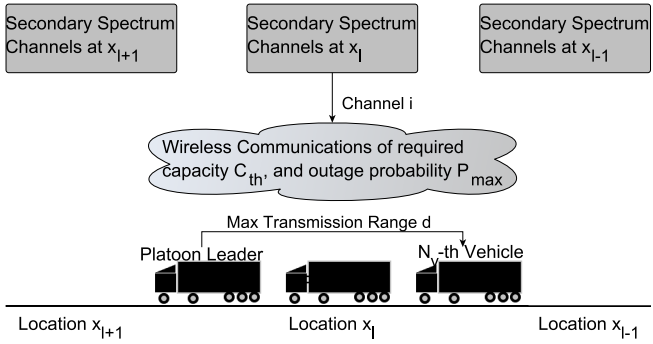}
\caption{ Illustration of the intra-platoon communication, with the use of the secondary spectrum channels.}
\label{fig:scenario_description}
\end{figure}

 In this paper, an autonomous platoon consisting of $N_\mathrm{v}$ vehicles is considered, as it is depicted in Fig.~\ref{fig:scenario_description}.  The platoon is claimed to travel over the route denoted as set of $\mathcal{L}$  geographical locations $\mathcal{X}=\{ \mathbf{x}_l\}_{l=1}^{\mathcal{L}}$, where $\mathbf{x}_l$ denotes vector of earth-centered earth-fixed (ECEF) geographical coordinates related to platoon location $l$. The route covers mostly highway areas. However, urban, and suburban areas can appear at the beginning and end of the route. The first vehicle is the platoon leader responsible for the management and behavior of the platoon. This functionality is done by sending proper information to the following $N_{\mathrm{v}}-1$ vehicles through a wireless channel. To ensure short inter-vehicle distance necessary for providing e.g. fuel saves, the wireless transmission must satisfy requirements on channel capacity $C_{\mathrm{th}}$, and reliability expressed e.g. as the maximum allowable probability of link capacity being below $C_{\mathrm{th}}$ denoted as $\mathcal{P}_{max}$. These requirements must be ensured for transmission between the platoon leader, and every other vehicle in the platoon. The resultant maximum transmission range is the distance between the platoon leader and $N_{\mathrm{v}}$-th vehicle, denoted as $d$. Although there is a dedicated frequency band for V2V communications around 5.9~GHz, the studies have shown that it will be highly occupied and transmission requirements for reliability would not be satisfied~\cite{sybis2018}. Thus, the intra-platoon communication is realized in terms of DSA, by utilization of one of the unoccupied secondary spectrum channels, denoted as channel~$i$. A major challenge is to choose a secondary spectrum channel $i$ from the set of available secondary spectrum channels $\mathcal{I}$ satisfying both $C_{\mathrm{th}}$, and $\mathcal{P}_{max}$ in each of the consecutive platoon locations $\mathcal{X}$. Although platoon can perform spectrum sensing, it is more beneficial to rely on past knowledge about the radio environment. It allows to e.g., characterize long term interference patterns over secondary spectrum channels $\mathcal{I}$ at given location $\mathbf{x}_l$, and plan future channel switching along the whole platoon route. We propose to create an intelligent database of location-dependent information about the radio environment i.e., REM. REM aims to assess the quality of secondary spectrum channels available in consecutive locations, based on past knowledge about interference distributions. The remaining part of this section will describe general REM architecture, algorithms for channel quality assessment and assignment, and method for reduction of REM size.  

\subsection{REM for Platoon Communication} \label{subsec:rem_4_platooning}

The general aim of the REMs is to provide the network with location-dependent data describing the radio environment, to, e.g., enable DSA. However, REM is not only the database. It provides intelligent mechanisms of location-dependent data acquisition, processing, and storage. The high-level REM architecture consists of the so-called Measurement Capable Devices (MCDs) providing the data, REM storage and acquisition unit, and REM manager responsible for proper data processing in order to handle communication with REM users~\cite{romero2015}.

\begin{figure}[!t]
\centering
\includegraphics[width=2.5in]{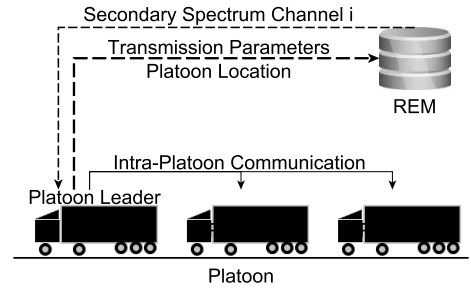}
\caption{Concept of REM utilization for platoon communications.}
\label{fig:rem_concept}
\end{figure}

For the purpose of intra-platoon communications, REM manager, storage, and acquisition units can be deployed as an extension to the existing cellular network infrastructure. Then, data exchange between the platoon leader and REM can be realized with the use of roadside units. The platoon can act both as an MCD, and REM user. When platoon acts as MCD it provides REM with batches of interference samples tagged with its geographical coordinates. These raw interference samples are processed by the REM manager unit to obtain their statistical models. Finally, model parameters, are saved in the REM storage unit. When platoon acts as a REM user it requests the REM manager to assign secondary spectrum channel~$i \in \mathcal{I}$ for intra-platoon communications at location $\mathbf{x}_l$ based on transmission parameters given by the platoon leader, and interference model parameters saved in REM. The high-level idea of this procedure is depicted in Fig.~\ref{fig:rem_concept}. The platoon leader sends to REM information about its location and desired transmission parameters. REM determines which secondary spectrum channel is the best for intra-platoon communication, and sends index~$i$ of this channel to the platoon leader. To enable optimization of secondary spectrum channel assignment along the whole platoon route, we consider centralized REM architecture. However, the major research issue related to the procedure of selection of secondary spectrum channel is the assessment of the available radio resources, possibly located at various frequency bands.   

\subsection{Outage Probability Estimation} \label{subsec:outage_probability_estimation}
 To assign a frequency band to a given platoon, a channel quality assessment method is necessary. The most effective approach from the perspective of platoon stability is to use the latency as a metric~\cite{zeng2019}. However, in general case its estimation is not straightforward, i.e.,  the latency depends on the used radio access technology and medium access protocol. Moreover to accurately estimate the latency the distribution of the incoming packets should be known. Instead, a reasonable comparison between radio channels, irrespective of the particular system properties, e.g., utilized coder or modulation, can be done on the basis of capacity being the upper bound estimate of the real system throughput. Such an approach is legitimated by the literature, e.g., to compare various frequency bands, in the context of Dynamic Spectrum Access~\cite{akylidiz2008}, or for the purpose of communications with unmanned aerial vehicles~\cite{zhang2019}. However, in Sec.~\ref{subsec:outage_and_latency}, we will show that there exists a relation between the radio channel capacity and latency. We decided to compute the capacity of a single channel as a sum over the capacities of component narrow frequency sub-channels. The first motivation is that various interference sources can have various frequency-specific emission characteristics. The second reason is that many contemporary systems utilize multi-carrier communication based
on the Orthogonal-Frequency-Division-Multiplexing (OFDM), e.g., 802.11p, Digital Video Broadcasting- Terrestrial (DVB-T), 5G. Thus Shannon capacity between most distanced cars within the platoon, at location $\mathbf{x}_l$,
can be computed as follows:
\begin{equation} \label{eq:wifi_capacity}
c^{(i, \mathbf{x}_l)} = \mathrm{B} \sum_{k \in \mathcal{K}} \log_{2}\left(1 + \frac{P_{tx}^{(i, \mathbf{x}_l)} \cdot L(d)^{(i, \mathbf{x}_l)}}{\sigma_{\mathrm{n}}^2 + I_{k }^{(i, \mathbf{x}_l)}}\right), 
\end{equation}
where, $i$ stands for the secondary spectrum channel index, $\mathcal{K}$ is the set of active sub-channels used in intra-platoon communications, $\mathrm{B}$ is the sub-channels spacing, $P_{tx}^{(i, \mathbf{x}_l)}$ is the transmitter power per sub-channel (maximum power reported by the platoon leader), $\sigma_{\mathrm{n}}^2$ is the noise power over sub-channel band, $I_{k}^{(i, \mathbf{x}_l)}$ stands for the interference power on sub-carrier $k$, and $L(d)^{(i,\mathbf{x}_l)}$ is distance $d$ dependent channel gain (including TX and RX antenna gains).
  Typically radio channel gain randomly varies over frequency, and time. There are several empirical models of this phenomenon proposed for the evaluation of V2V communications~\cite{marum2007}.  However, measurement studies have shown that when 
 vehicles follow each other at close distance, e.g., below 200~m, there would be mainly Line-Of-Sight (LOS) propagation between them~\cite{nilsson2017}.  While there exists an advanced radio channel model, utilizing geometry of the surrounding scatters~\cite{boban2014}, this will be not useful for this application. The platoon would not have such detailed information about e.g., the location of buildings and other vehicles. Thus, less complex, yet well-established distance-dependent models can be used,e.g., two slope pathloss model~\cite{cheng2007}.
 Although channel impulse response will influence a single sub-channel, its influence should be averaged when considering the whole band. Moreover, the channel response variations caused by fast fading cannot be estimated before transmission occurs in the channel.     
 On the other hand, the interference $I_{k}^{(i, \mathbf{x}_l)}$ from primary systems can significantly vary over locations and frequencies affecting capacity. 
 As a result, we propose to make an assumption about the flat channel, to focus on the interference impact on the channel capacity.
 

Crucial for the platooning purpose is the reliable, transmission of the messages containing information about platoon leader velocity and acceleration. The low communication reliability and high latency can significantly decrease the benefits from platoon driving pattern, or in the worst cause a platoon crash. However, the required capacity is relatively low, e.g., using 802.11p safety messages are transmitted using minimal supported bit-rate~\cite{campolo2013}. However, the platoon requires that the capacity is available, e.g., it is more important to provide often the minimum required channel capacity, than high instantaneous channel capacity rarely. As such channel outage probability is of higher importance than ergodic or maximal channel capacity. Due to the randomly changing interference, the capacity $c^{(i, \mathbf{x}_l)}$ becomes a random variable. As a result the probability of capacity being below a certain threshold
is of high importance. Such situation occurs when the temporal signal-to-interference ratio (SINR) is very low. Thus it is reasonable to use the Shannon formula simplification for low SINR~\cite{tse2005}: $\log_{2}\left(1 + \frac{P_{tx}^{(i, \mathbf{x}_l)} \cdot L(d)^{(i, \mathbf{x}_l)}}{\sigma_{\mathrm{n}}^2 + I_{k }^{(i, \mathbf{x}_l)}}\right) \approx \frac{1}{\ln{2}}\cdot\left( \frac{P_{tx}^{(i, \mathbf{x}_l)} \cdot L(d)^{(i, \mathbf{x}_l)}}{\sigma_{\mathrm{n}}^2 + I_{k }^{(i, \mathbf{x}_l)}}\right)$.
This allows for the following transformation of~\eqref{eq:wifi_capacity}: 
\begin{equation} \label{eq:wifi_low_sinr}
    c^{(i, \mathbf{x}_l)} \approx \frac{\mathrm{B} \cdot P_{tx}^{(i, \mathbf{x}_l)} \cdot L(d)^{(i, \mathbf{x}_l)}}{\ln{2}} \cdot \sum_{k \in \mathcal{K}} \frac{1}{\hat{I}_k^{(i, \mathbf{x}_l)}},
\end{equation}
where $\hat{I}_k^{(i,\mathbf{x}_l)}=\sigma_{\mathrm{n}}^2 + I_{k}^{(i, \mathbf{x}_l)}$. Now the outage probability formula is given as:
\begin{equation} \label{eq:outage_prob}
    \mathcal{P}_{out}^{(i, \mathbf{x}_l)} = \mathcal{P}\left(\sum_{k \in \mathcal{K}} \frac{1}{\hat{I}_k^{(i, \mathbf{x}_l)}} < \frac{\ln{2}\cdot C_{\mathrm{th}}}{\mathrm{B} \cdot P_{tx}^{(i, \mathbf{x}_l)} \cdot L(d)^{(i,\mathbf{x}_l)}}\right),
\end{equation}
where $C_{\mathrm{th}}$ is the minimum required link capacity that supports the safe operation of the platoon. It can bee seen that due to the Shannon formula simplification for low SINR, the outage probability is only a function of single random variable, i.e., cumulative interference power $\sum_{k \in \mathcal{K}} \frac{1}{\hat{I}_k^{(i, \mathbf{x}_l)}}$. Moreover, the interference distribution in a particular location can be characterized in terms of one instead of $\mathcal{K}$-dimensional distribution. This will much simplify the data representation in REM.
The interference in wireless communications usually follows the log-normal distribution~\cite{cardieri2001}. Following this phenomenon, it seems reasonable to take the logarithm of~\eqref{eq:outage_prob}. As a result, we expect to deal with Gaussian-like distributions instead of log-normal, which are much more complex to model.
Both sides of~\eqref{eq:outage_prob} are always positive, thus logarithm of both sides can be found as: 
\begin{equation} \label{eq:outage_prob2}
    \mathcal{P}\left( \chi^{(i, \mathbf{x}_l)} < \ln{ \frac{\ln{2}\cdot C_{\mathrm{th}}}{\mathrm{B} \cdot P_{tx}^{(i,\mathbf{x}_l)} \cdot L(d)^{(i,\mathbf{x}_l)}}}\right),
\end{equation}
where $\chi^{(i, \mathbf{x}_l)} = \ln{\sum_{k \in \mathcal{K}} \frac{1}{\hat{I}_k^{(i, \mathbf{x}_l)}}}$ is the interference power distribution related to the frequency channel~$i$, and location $\mathbf{x}_l$. Now the outage probability, can be computed, using the distribution of $\chi^{(i, \mathbf{x}_l)}$, which characterizes interference, and some constants provided by the platoon leader, i.e., its transceiver bandwidth, maximal transmit power and minimum required link capacity. The $\chi^{(i, \mathbf{x}_l)}$ distribution parameters related to geographical locations are envisioned to be stored in REM. 
The modeling of $\chi^{(i, \mathbf{x}_l)}$ 
on the basis of field measurements 
will be described in Sec.~\ref{subsec:data_analysis}. 

\subsection{ Relation Between Outage Probability and Latency} \label{subsec:outage_and_latency}
Although in this paper the outage probability is considered as a MAC-independent channel quality assessment metric, it can be mapped to the transmission latency, if necessary. Since there is usually no acknowledgment procedure in V2V communications when sending safety messages, and some Listen-before-talk procedure is used (e.g., in IEEE 802.11p standard), it can be expected that a delay occurs when the channel is busy. In the idealized model the channel is busy if its instantaneous capacity is lower than the capacity required for the transmission $C_{\mathrm{th}}$. At such time instance the achieved data rate equals zero. On the other hand, if the channel capacity exceeds required threshold, the data rate is fixed as a result of a single modulation-coding scheme used for broadcast messages and equals for the perfect system $C_{\mathrm{th}}$. This allows to calculate mean data rate in a channel as $(1-\mathcal{P}(c^{(i,\mathbf{x}_l )}<C_{\mathrm{th}})) C_{\mathrm{th}}$. This is the upper bound of the mean data rate that can be achieved in this channel. From this perspective, we can formulate the lower bound of the latency as a function of outage probability, and under the assumption of fixed packet size $D$. The formula is given by:
\begin{equation} \label{eq:latency}
    \tau^{(i, \mathbf{x}_l)} \geq \frac{D}{(1-\mathcal{P}(c^{(i,\mathbf{x}_l )}<C_{\mathrm{th}})) C_{\mathrm{th}}}.
\end{equation}
A similar way of modeling latency is used in some works focused on the control algorithms, e.g.,~\cite{zeng2021federated, zeng2019}. Observe that the actual mean delay $\tau^{(i, \mathbf{x}_l)}$ will be probably higher as a result of radio access technology not achieving channel capacity and because of limitations of the Medium Access Control scheme. We propose to observe the outage probability, a well-established metric, irrespective of various possible transmission technologies of the platoon and the neighboring wireless systems.

\subsection{Radio Channel Assignment} \label{subsec:radio_channel_assignment}

\begin{figure}[!t]
\centering
\includegraphics[width=4.2in]{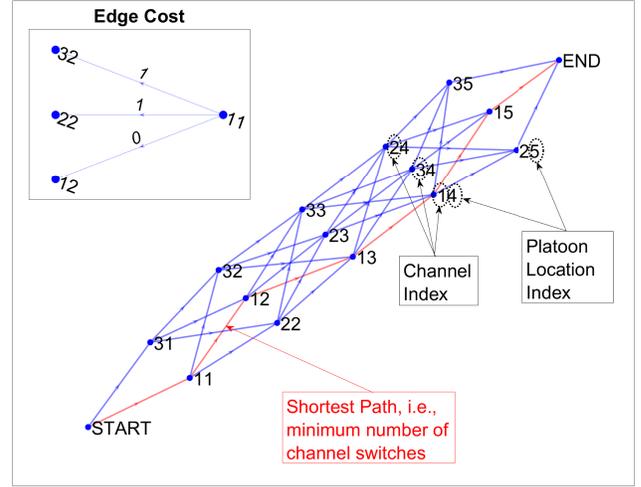}
\caption{ Example of a graph representation of channel switching problem. There are 3 channels, and 5 platoon locations. Nodes represent available radio channels in the secondary spectrum.}
\label{fig:channel_assignment}
\end{figure}

Using \eqref{eq:outage_prob2} the outage probability related to every channel $i$ in each location $\mathbf{x}_l$ belonging to the platoon route can be estimated.
The most straightforward channel selection assumes selection of the channel related to the lowest outage probability:
\begin{equation} \label{eq:optimal_channel_assignment}
    \hat{i}^{(\mathbf{x}_l)} = \arg \min_i{\mathcal{P}\left( \chi^{(i, \mathbf{x}_l)} < \ln{ \frac{\ln{2}\cdot C_{\mathrm{th}}}{\mathrm{B} \cdot P_{tx}^{(i,\mathbf{x}_l)} \cdot L(d)^{(i,\mathbf{x}_l)}}}\right)}.
\end{equation}
However, such an approach can result in frequent channel switching. In the worst-case channel will be switched in every new location. This is not encouraged from the perspective of connection stability. In practical implementations, it is enough to ensure the outage probability being below the given threshold $\mathcal{P}_{\mathrm{max}}$. 
Thus we can define a channel assignment as the optimization problem, where the target is to minimize the number of channel switches along the platoon route, subject to outage probability being below the required value: 
\begin{equation} \label{eq:minimizaton_of_switches}
\begin{aligned}
&  \min_{i^{(\mathbf{x}_r)}} &&{\sum_{l=1}^{\mathcal{L}-1} s(i^{(\mathbf{x}_l)}, i^{(\mathbf{x}_{l+1})})}, \\
&    s.t. && \mathcal{P}(c^{(i^{(\mathbf{x}_l)}, \mathbf{x}_l)} < C_{\mathrm{th}}) \leq \mathcal{P}_{\mathrm{max}}
\end{aligned}
\end{equation}
 where $l$ is the index representing consecutive platoon locations, and $\mathcal{L}$ is the total number of locations along the platoon route,
and $s(i^{(\mathbf{x}_r)}, i^{(\mathbf{x}_{r+1})})$ is the cost of a single channel switch, defined as follows:
\begin{equation} \label{eq:cost_function}
    s(i^{(\mathbf{x}_l)}, i^{(\mathbf{x}_{l+1})}) = \begin{cases} 1, \text{if, } i^{(\mathbf{x}_l)} \neq i^{(\mathbf{x}_{l+1})}\\
    0, \text{if } i^{(\mathbf{x}_l)} = i^{(\mathbf{x}_{l+1})}
    \end{cases}.
\end{equation}
 We propose to solve the optimization problem~\eqref{eq:minimizaton_of_switches} by using graph theorem. The graph representation of channel assignment is depicted in Fig.~\ref{fig:channel_assignment}. The example graph is constructed assuming 3 channels and 5 locations. Nodes represent radio channels providing sufficient outage probability, in consecutive locations. Graph edges stand for the channel switching cost. Following~\eqref{eq:cost_function}, the cost is equal to 1, when channel switching occurs, and 0 otherwise. Additionally, two nodes "START", and "END" are introduced with edge costs equal to 0. The optimization problem can be now solved in terms of finding the shortest path between "START" and "END" nodes. Because all graph edges are non-negative, the optimal solution to the problem of finding the shortest path can be computed using Dijkstra algorithm~\cite{Mehlhorn2008}. The problem can be alternatively expressed as a linear programming problem. While linear programming belongs to a class of convex optimization, the Dijkstra algorithm provides global optimum~\cite{boyd2004convex}. 
 
 The worst-case computational complexity of the Dijkstra algorithm on any directed graph is~\cite{barbehen1998}:
 \begin{equation}
     O(E + V\log V),
 \end{equation}
 where $E$ denotes the total number of edges and $V$ the number of vertices. The worst-case number of vertices is calculated considering that in each location, out of $\mathcal{L}$ along the platoon route, there are $\mathcal{I}$ possible wireless channels. While each of these location-channel pairs constitutes a vertex, the total number of vertices can be calculated as:
\begin{equation}
   V=\mathcal{L} \cdot \mathcal{I} 
\end{equation}
 From each of these vertices, $\mathcal{I}$ edges start, being connected to the maximum of $\mathcal{I}$ possible channels in the next location. The worst-case number of edges is the number of vertices $V$ multiplied by the maximum number of available radio channels $\mathcal{I}$ giving:
 \begin{equation}
     E = V\cdot \mathcal{I} = \mathcal{L}\cdot \mathcal{I}^2.
 \end{equation}
The resultant worst-case computational complexity of the Dijkstra algorithm is given by:
\begin{equation}
    O(E+V \log V) = O\left( \mathcal{L}\cdot \mathcal{I}^2 + \mathcal{L}\cdot \mathcal{I} \log {\mathcal{L}\cdot \mathcal{I} } 
    \right).
\end{equation}
 
 The channel assignment scheme computed in REM with the use of Dijkstra algorithm can be sent to the platoon leader. If the platoon route changes or there is a significant interference distribution change, the calculations have to be repeated.

\subsection{Reduction of REM Size} \label{subsec:reduction_rem_size}

One could imagine that there will be potentially a lot of data, obtained over many locations in REM. On the other hand, there is a possibility that there exists a spatial correlation between them, i.e., the interference $\chi^{(i, \mathbf{x}_l)}$ follows the same distribution over neighbouring geographical locations. The challenge is to choose a clustering algorithm proper for REM data. Successful clustering will be able to increase the statistical correctness of the interference distribution estimates and decrease the REM size, i.e., the amount of data that has to be processed. It should be able to deal both with distance in terms of distribution, and geographical position, i.e., the aim is to connect geographically close points, where $\chi^{(i, \mathbf{x}_l)}$ follows the same distribution. One simple and popular clustering algorithm is the so-called K-means~\cite{vassil2007}. The major drawback of this method is that it requires a-priori the number of clusters to be created. A method that overcomes this issue is, e.g., hierarchical clustering~\cite{johnson1967hierarchical}. However, this method cannot simultaneously deal with two distance measures. The solution, where both numbers of clusters is not fixed, and there is a possibility of incorporating both geographical distance and distribution similarity measure is the so-called Density-Based Spatial Clustering of Applications with Noise (DBSCAN)~\cite{ester1996}. Originally DBSCAN, has two parameters: neighbourhood radius $\epsilon$, defining maximum radius used for neighbourhood search, and $\mathrm{N_{p}}$ being the minimum number of points to formulate a cluster. However, there is a possibility to define a more complex function for neighbourhood determination. In order to cluster REM entries being both in geographically close distance, and characterized with similar $\chi^{(i,\mathbf{x}_l)}$ distributions, we propose to define neighbourhood radius for geographical distance $\epsilon_{g}$, and neighbourhood radius for distribution similarity $\epsilon_{KS}$. The geographical distance between two points is computed in terms of Euclidean distance, which is proper when using ECEF coordinates: 
\begin{equation} \label{eq:geographical_distance}
    d_g(\mathbf{x}_l, \mathbf{x}_{l^\prime}) = ||\mathbf{x}_l - \mathbf{x}_{l^\prime}||,
\end{equation}
where $||\cdot||$ denotes an Euclidean norm, and $\mathbf{x}_l$ vector of ECEF geographical coordinates related to REM entries $l$, and $l^\prime$ respectively. The neighborhood radius for geographical distance $\epsilon_g$ must be obtained based on empirical studies. The similarity measure between interference distributions is defined in terms of Kologomorov-Smirnoff distance~\cite{massey1951}:
\begin{equation} \label{eq:ks_distance}
d_{\mathrm{KS}}(i,\mathbf{x}_l,\mathbf{x}_{l^\prime}) = \sup_z |F(z)^{(i,\mathbf{x}_l)} - F(z)^{(i,\mathbf{x}_{l^\prime})}|,
\end{equation}
where $F(z)^{(i,\mathbf{x}_l)}$ denotes the cumulative distribution function of $\chi^{(i, \mathbf{x}_l)}_i$. The neighbourhood radius can be computed on the basis of critical value $c(\alpha)$ related to the significance level $\alpha$:
\begin{equation} \label{eq:ks_test_threshold}
    \epsilon_{KS} = c(\alpha) \cdot \sqrt{\frac{n+m}{n \cdot m}},
\end{equation}
where $n$ and $m$ are the number of data samples used to estimate $F(z)^{(i,\mathbf{x}_l)}$ and $F(z)^{(i,\mathbf{x}_{l^\prime})}$,respectively.  Precomputed values of $c(\alpha)$, proper for Kolmogorov distribution can be found in tables~\cite{CALIXTO20161}. Let us introduce two auxiliary logic formulas: $Q_g(\mathbf{x}_l, \mathbf{x}_{l^\prime}): d_g(\mathbf{x}_l, \mathbf{x}_{l^\prime}) < \epsilon_g$, and $Q_{KS}(\mathbf{x}_l, \mathbf{x}_{l^\prime}): \forall \,i, \; d_{\mathrm{KS}}(i,\mathbf{x}_l, \mathbf{x}_{l^\prime}) < \epsilon_{KS}$. 
Finally, the function that determines if two data points in REM can be classified as neighbors during DBSCAN procedure
can be formulated as: 
\begin{equation} \label{eq:dbscan_nieghbourhood_func}
    f(\mathbf{x}_l, \mathbf{x}_{l^\prime}) = \begin{cases}
    1, \text{ if } Q_g(\mathbf{x}_l, \mathbf{x}_{l^\prime}) \land Q_{KS}(\mathbf{x}_l, \mathbf{x}_{l^\prime})  \\
    0, \text{ otherwise}.
    \end{cases}
\end{equation}
Using the above equation as an alternative distance function, DBSCAN can be performed to cluster the REM data. The $\epsilon_{g}$ must be obtained based on empirical study. It is reasonable to start from a low $\epsilon_g$ value and increase it and observe, e.g., maximum distance to the nearest neighborhood over all clusters. The REM entries within a single cluster share similar $\chi^{(i,\mathbf{x}_l)}$ distribution over all channels $i$, and are geographically close to each other. These REM entries can be merged, to reduce REM size. One important property of DBSCAN is that a particular group of points, i.e., REM entries, can remain without cluster assignment. REM entry without cluster assignment cannot be merged to another, because it is either characterized with unique $\chi^{(i,\mathbf{x}_l)}$ distribution, or too geographically distanced from other REM entries.

\section{Measurement Campaign} \label{sec:measurement_campaign}

\begin{figure}[!t]
\centering
\includegraphics[width=4.1in]{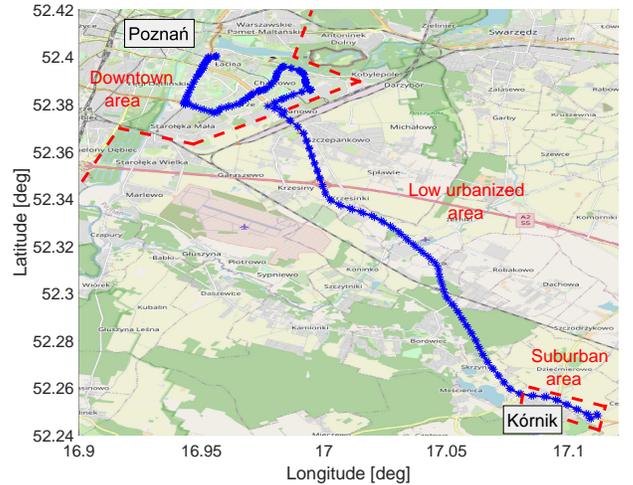}
\caption{Measurement campaign route between Poznań, and Kórnik, mostly along the S11 speedway.}
\label{fig:route_map}
\end{figure}

To assess the proposed methods with realistic input data field measurements were conducted. We have arbitrarily selected three non-overlapping WiFi\texttrademark{}  channels of center frequencies $2.412$~GHz, $2.437$~GHz, and $2.462$~GHz, i.e., channel numbered widely as 1, 6, and 11. The 2.4 GHz band is treated as the \emph{worst case} scenario with a high number of interference sources of various types.
The data has been collected along the route between Poznań, and Kórnik in Poland, as depicted in Fig.~\ref{fig:route_map}. The route has been traveled from Poznań to Kórnik and back. The major part of the route was a two-lane speedway S11 placed mostly in the low urbanized area. However, the few first kilometers of the measurement campaign are carried in Poznan downtown. The last part of the measurements is carried close to Kórnik representing a small town or suburban scenario. As such the measurements should be representative of all main types of propagation environments.   

\subsection{Measurement Setup}

The measurement data were collected in terms of in-phase and quadrature (IQ) received signal samples collected using the Rhode\&Schwarz FSL6 spectrum analyzer. There was an omnidirectional antenna PCTEL LPBMLPVMB/LTE of $3$ dBi gain, installed on the rooftop of the car, and attached to the spectrum analyzer. The spectrum analyzer was configured to obtain maximal sensitivity (turned on preamplifier and 0 dB input attenuation) resulting in about -84 dBm of thermal noise (measured over $20$~MHz bandwidth) reported with the sampling frequency of 20 Msps. Also, there was a U-Blox GPS receiver reporting car position every second with the use of messages defined by the National Marine Electronics Association (NMEA) protocol. Both GPS module and spectrum analyzer are plugged into the laptop using USB, and Ethernet connections, respectively. The whole setup was controlled by the Matlab software, including data capturing, and time stamping both the IQ samples from spectrum analyzer and position from GPS receiver. The power supply was provided from the car 12~V socket.    

The spectrum analyzer collects 65536 IQ samples in one frequency channel at one measurement. Next, the center frequency is sequentially switched between the chosen WiFi channels: 1st, 6th, and 11th. Each IQ sample vector is tagged with a geographical position and time. In most cases, the sensing in each channel is repeated approximately every 250 ms. However, after 50 observations the laptop saves the observed samples as a file to the hard drive. This introduces around 6.5 seconds of stoppage in the data collection. There were 248 files with measurement data captured in total along the route.     

\subsection{Data Analysis} \label{subsec:data_analysis}
\begin{figure}[!t]
\centering
\includegraphics[width=3.5in]{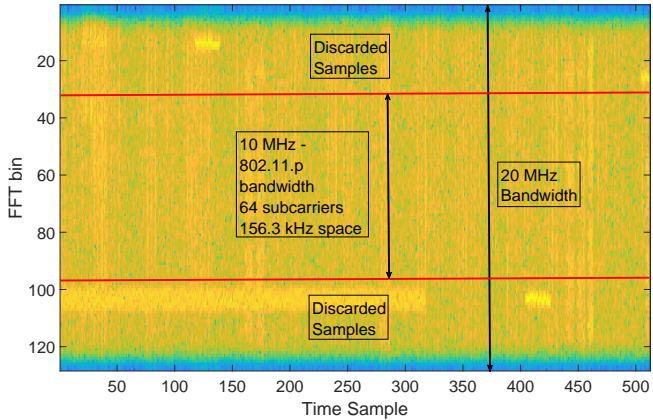}
\caption{Example of a spectrogram for the received signal. Samples at both frequency edges are discarded to keep coherence with 802.11p bandwidth and to not consider distortions caused by spectrum analyzer frequency characteristics.}
\label{fig:setup}
\end{figure}

To extract the interference and noise power ($\hat{I}_k^{(i,\mathbf{x}_l)}$) related to each of the sub-carriers a Discrete Fourier Transform (DFT) was applied to the collected IQ samples. While the samples were collected with a sampling rate of 20 MHz, DFT of 128 points was used. This results in the same subcarrier spacing as in the IEEE 802.11p standard. As IEEE 802.11p uses 10 MHz of bandwidth with 64 subcarriers, only the middle 64 subcarriers are considered for further processing. This allows discarding subcarriers belonging to a roll-off frequency range of the spectrum analyzer. An example of a spectrogram with the bandwidth of interest marked is depicted in Fig.~\ref{fig:setup}. 
The measurement data processed with the use of the 128 point DFT is available online~\cite{measurement_samples}.

\begin{figure}[!t]
\centering
\includegraphics[width=3.6in]{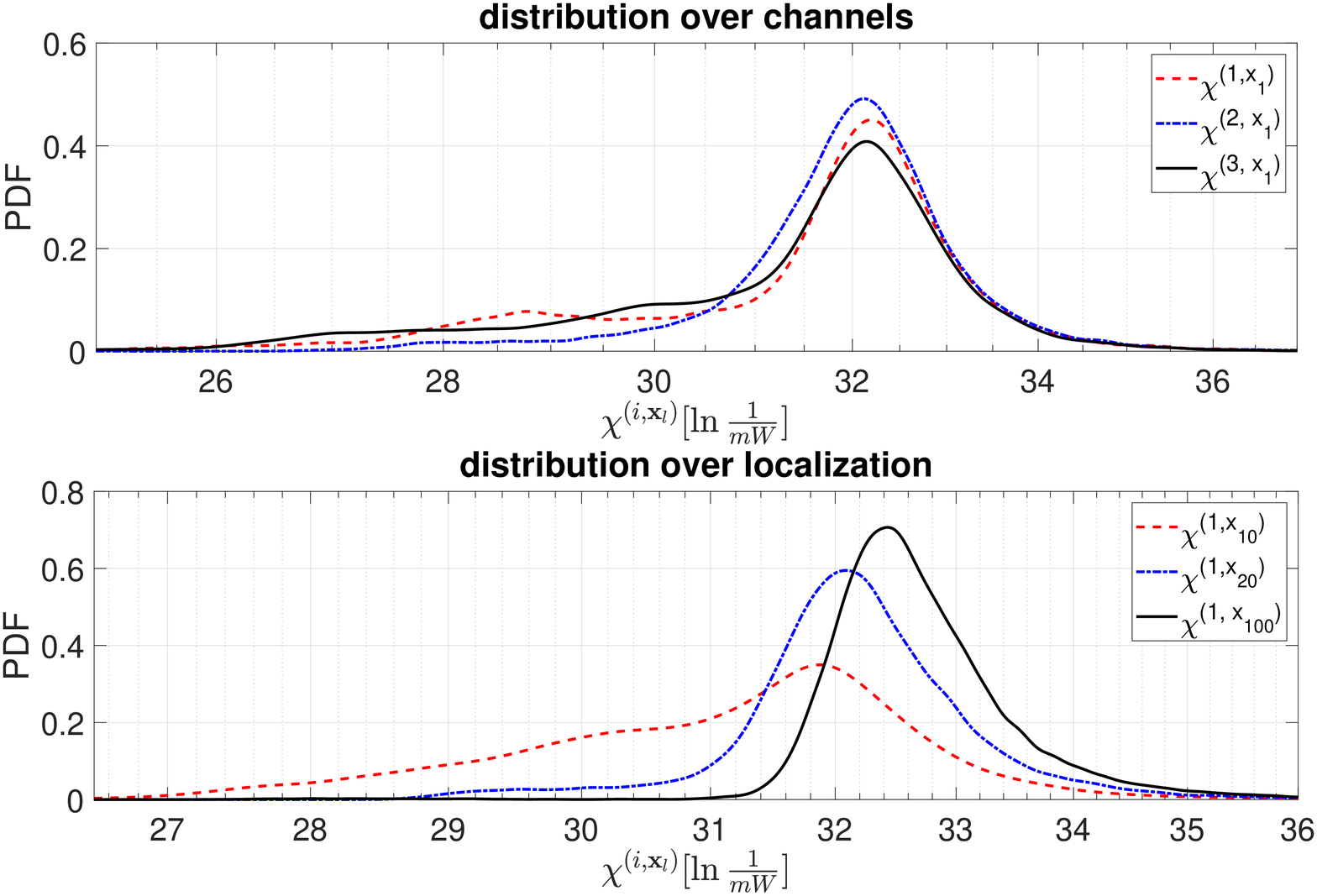}
\caption{Probability Density Function of $\chi^{(i, \mathbf{x}_l)}$ over three WiFi channels in a single location, and over a single WiFi channel in three locations.}
\label{fig:interf_distribution}
\end{figure}

As was already mentioned captured data is organized in files, containing vectors of IQ samples. These vectors of IQ samples can be seen as the batches of data measured by the platoon at a certain location and send to REM for further processing.
Measurement data having the same geographical position tag reported via NMEA messages are used to formulate one REM entry. 
Examples of the estimated probability density functions (PDFs) of $\chi^{(i, \mathbf{x}_l)}$ over various WiFi channels and locations are depicted in Fig.~\ref{fig:interf_distribution}. The first observation is that the distributions vary over channels, thus outage probability also will vary. As a result, there would exist a channel being the best for opportunistic transmission in a particular location. Second observation is that the $\chi^{(i, \mathbf{x}_l)}$ distribution varies over locations. This justifies the idea of REM and the spatial divisions of data. Final observation is that $\chi^{(i, \mathbf{x}_l)}$ follow non trivial distributions. In some cases these have multimodal distributions, i.e., having more than a single peak. In other cases $\chi^{(i, \mathbf{x}_l)}$ distribution is characterized with non-regular tails.
This can be justified by the nature of interference generated from many sources with varying characteristics. If there was only a single interference source, there would be only one peak in PDF of $\chi^{(i, \mathbf{x}_l)}$. However, if there are several sources of temporal interference having different transmission characteristics, then multiple peaks are expected to occur in $\chi^{(i, \mathbf{x}_l)}$ PDF, representing different configurations of interference sources. Under some channels, and locations there might be one most common configuration of interference sources, then instead of clearly visible multiple peaks, the distribution tails are shaped in a non-trivial way. This is expected in the 2.4 GHz band where many communication systems operate simultaneously, e.g., WiFi and Bluetooth.  Such a non-trivial, e.g., multimodal distributions can be efficiently modeled using the so-called Gaussian Mixture Model (GMM)~\cite{bishop2006}. 
GMM represents the random variable PDF as the weighted sum of $J$ component Gaussian densities. In our case modeling of the one-dimensional distribution of $\chi^{(i, \mathbf{x}_l)}$ is considered: 
\begin{equation}\label{eq:gmm}
    p(\chi^{(i, \mathbf{x}_l)}) = \sum_{j=1}^{J}w_j \mathcal{N}(\chi^{(i, \mathbf{x}_l)} |\mu_j, \sigma_j),
\end{equation}
where $p(\chi^{(i, \mathbf{x}_l)})$ denotes the distribution of $\chi^{(i, \mathbf{x}_l)}$, $w_j$ is the $j$-th mixture component weight, i.e., the probability that $\chi^{(i, \mathbf{x}_l)}$ comes from the $j$-th mixture component, $\mathcal{N}(\chi^{(i, \mathbf{x}_l)} |\mu_j, \sigma_j)$ is the conditional Gaussian distribution of $\chi^{(i, \mathbf{x}_l)}$, i.e., Gaussian distribution of $\chi^{(i, \mathbf{x}_l)}$, under assumption that it comes from the $j$-th mixture component. Finally $\mu_j$ is the mean, and $\sigma_j$ is the standard deviation of $j$-th mixture component. 
GMM can be then represented in this case by the vector of weights $\mathbf{w}$, the vector of components means $\mathbf{\mu}$, and vector of standard deviations $\mathbf{\sigma}$. Such a model seems adequate for being stored in REM, as it represents the distribution with only a few parameters instead of, e.g., histogram, direct PDF, or raw interference samples. 
A common approach to obtain estimates of distribution parameters is the so-called maximum likelihood (ML) estimation. However, in the case of GMM the log-likelihood function is nonlinear, i.e., the logarithm of the sum over exponential functions~\cite{bishop2006}. There is no closed-form ML GMM parameter estimator. However, there is an iterative algorithm called expectation-maximization (EM), which can be utilized~\cite{dempster1977}. 

\begin{figure}[!t]
\centering
\includegraphics[width=3.7in]{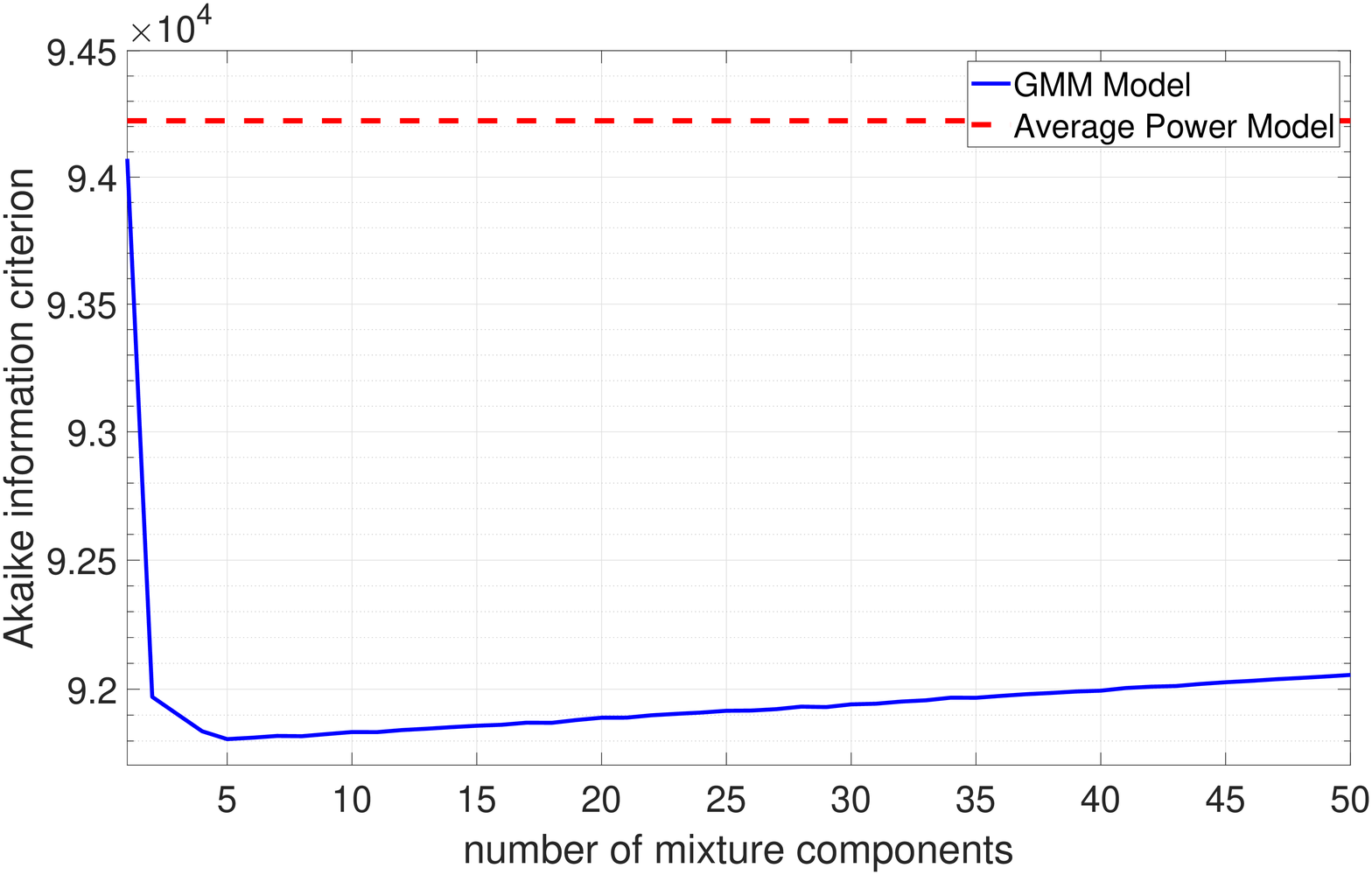}
\caption{Example of AIC related to fitting GMM to measurement data, while varying number of components $J$.}
\label{fig:GMM_fit}
\end{figure}

One of the challenges related to the modeling distribution with GMM is obtaining the proper number of components, i.e., high enough to reflect accurately interference distribution but low enough not to introduce overfitting. 
To ensure this, GMM fits are evaluated in terms of Akaike Information Criterion (AIC), which is the function of model log-likelihood, and the number of parameters~\cite{akaike1974}:
\begin{equation} \label{eq:aic}
    \mathrm{AIC}_J = \omega_J - 2\ln{\hat{L}},
\end{equation} where $\omega_J$ is the number of parameters of GMM with $J$ components, and $\hat{L}$ is the maximum of likelihood function related to GMM having $J$ components. In the case of GMM modeling one-dimensional distribution of $\chi^{(i,\mathbf{x}_l)}$, $\omega_J$ equals $3J$. We have fitted GMM to measurement data, using various number of mixture components $J$. For each model the corresponding $\mathrm{AIC}_J$ was computed. In addition AIC related to the model exploiting only average value of $\chi^{(i,\mathbf{x}_l)}$ is computed. This model stands for the state-of-the-art approach in REM~\cite{katagiri2018, sybis2018,sroka2020}. A representative result, obtained by fitting interference distribution from a single measured samples file, is depicted in Fig.~\ref{fig:GMM_fit}. The lower $\mathrm{AIC}_J$ value is the better, i.e., the best balance between GMM accuracy and number of parameters is provided.
It can be seen that GMM model significantly outperforms the state-of-the-art average power model for any considered number of mixture components.
Between the GMM models, the one consisting of 5 components offers the lowest value of AIC. 
Based on these measurements we decided to model $\chi^{(i, \mathbf{x}_l)}$ as 5 component GMM for all REM entries 
The obtained GMM parameters would be then stored in REM, for the purpose of channel assignment (see Sec.~\ref{subsec:radio_channel_assignment}). 

 In fully deployed REM we can expect that newly collected interference samples should result in the update of the GMM parameters stored in REM, e.g., newer samples are more important than the old ones as the interference characteristics can evolve in time. This can affect both the number of mixture components $J$, and mixture component parameters: $w_j$, $\mu_j$, and $\sigma_j$. One method for adaptive adjusting number of mixture components, and recursive update of $w_j$, $\mu_j$, and $\sigma_j$, utilizing exponential averaging had been described in~\cite{zivkovic2004}. However, application and in-depth verification of this approach would require a much larger data set. We leave it open for future research.    

\subsection{DBSCAN Clustering} \label{subsec:clustering_results}
 The REM constructed, according to the interference modeling procedure described in the previous subsection is claimed to contain some redundant data. I.e., some of the GMMs parameters describing $\chi^{(i, \mathbf{x}_l)}$ are expected to be very similar. To reduce those similarities  REM size reduction algorithm using DBSCAN (see Sec.~\ref{subsec:reduction_rem_size})  is launched on the REM.  

\begin{figure}[!t]
\centering
\includegraphics[width=4.0in]{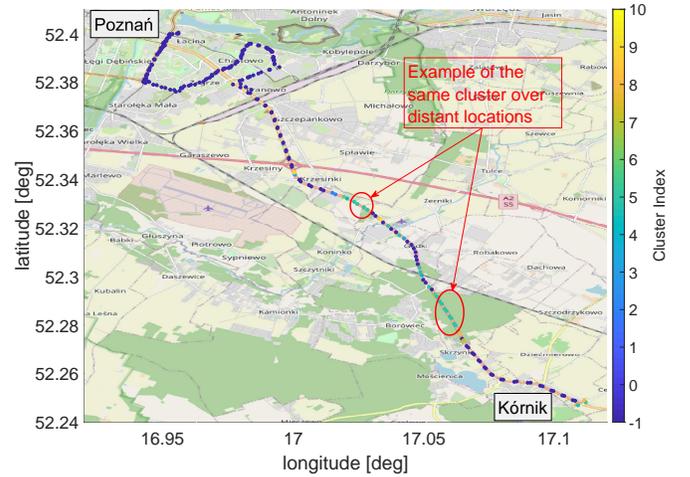}
\caption{Results of REM entries clustering using DBCAN, for $\mathrm{N_p = 2}$, $\epsilon_{\mathrm{KS}}=0.12$, and $\epsilon_{\mathrm{g}}= \infty$. The points marked with index $-1$ could not be clustered. 
}
\label{fig:db_cluster1}
\end{figure}

\begin{figure}[!t]
\centering
\includegraphics[width=3.7in]{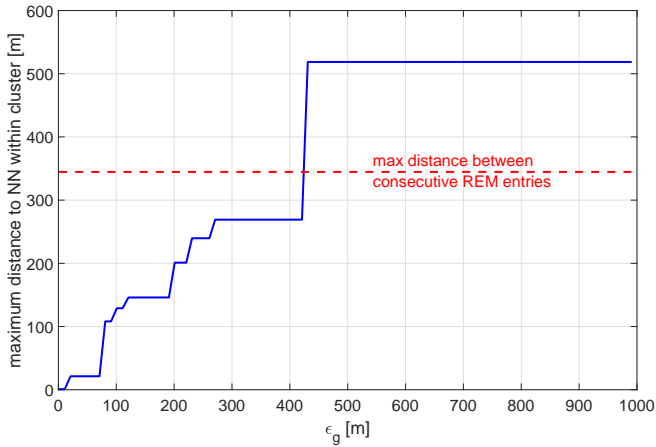}
\caption{Dependency between the geographical neighborhood radius $\epsilon_g$, and maximum NN over all clusters. Neighbourhood radius for distribution similarity is $\epsilon_{\mathrm{KS}}=0.012$. }
\label{fig:max_knn_cluster}
\end{figure}

\begin{figure}[!t]
\centering
\includegraphics[width=3.9in]{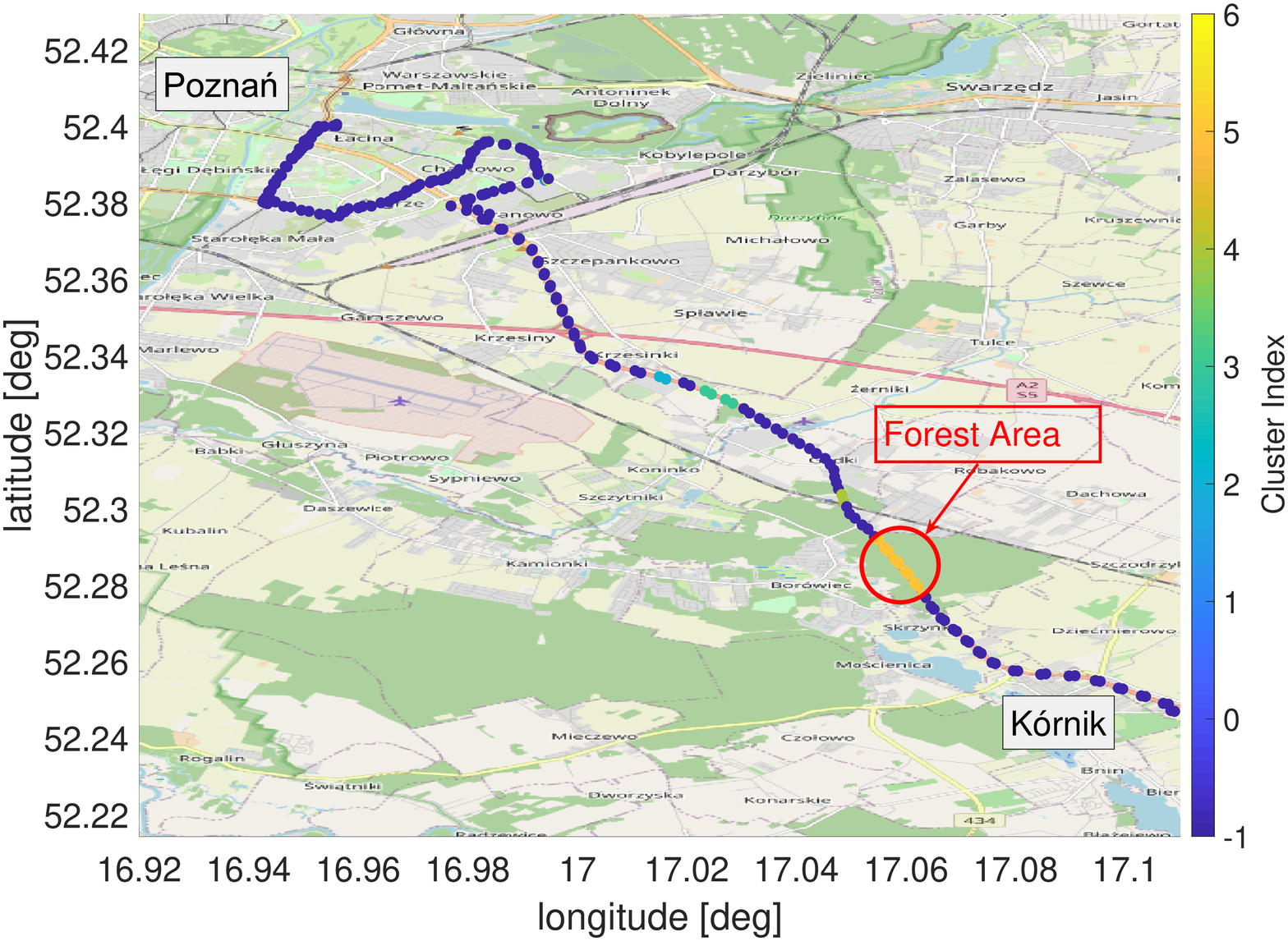}
\caption{Results of REM entries clustering using DBCAN, for $\mathrm{N_p = 2}$, $\epsilon_{\mathrm{KS}}=0.12$, and $\epsilon_{\mathrm{g}}= 400$~m. The points marked with index $-1$ could not be clustered. }
\label{fig:db_cluster2}
\end{figure}

The DBSCAN-based algorithm takes three arguments: $N_\mathrm{p}$ minimum number of points to formulate cluster, neighbourhood radius for geographical distance $\epsilon_{\mathrm{g}}$ and distribution similarity $\epsilon_{\mathrm{KS}}$. We set $N_\mathrm{p}=2$ 
, because it is the lowest supported value, which results in the highest opportunity for finding similar REM entries. The value of $\epsilon_{\mathrm{KS}}$ is computed using~\eqref{eq:ks_test_threshold}. Each distribution was obtained using the same number of samples: $n=m=25600$, i.e., there were 50 vectors of measurement data utilized to create a single REM entry, after 128-point DFT each vector has a time resolution of $512$ samples. 
We took commonly used in statistic $c(\alpha = 0.05) = 1.358$. The resultant neighbourhood radius for distribution similarity is $\epsilon_{\mathrm{KS}} = 0.012$. Initially the neighbourhood radius for geographical distance would be set to be very large in order to evaluate only similarities between distributions. The result of DBSCAN performed on the REM data is depicted in Fig.~\ref{fig:db_cluster1}. 
There, are 10 clusters created, that are allowed to achieve about 12~\% reduction of the REM size.
It can be seen that more REM entries can be clustered together in the low urbanized area. However, there are REM entries having the same $\chi^{(i, \mathbf{x}_l)}$ distributions but quite distant to each other, e.g., the maximum distance to the nearest neighbor (NN) within the same cluster is about $5$~km. This is a negative phenomenon in terms of REM entries merging as the intuition is that more distant points should create separate areas. This justifies the introduction of the second neighborhood radius $\epsilon_{\mathrm{g}}$. We have examined several $\epsilon_{\mathrm{g}}$ values. The resultant maximum distance to NN within-cluster is depicted in Fig.~\ref{fig:max_knn_cluster}. Results are compared to the maximum distance between consecutive REM entries marked with the dashed line. It can be observed that slightly above the threshold of $400$~m, the maximum distance to NN within-cluster exceeds the maximum distance between original REM entries. The geographical neighborhood radius $\epsilon_{\mathrm{g}}$ is set to the value of $400$~m. The results of DBSCAN clustering obtained for $\epsilon_{\mathrm{g}}=400$~m, are depicted in Fig.~\ref{fig:db_cluster2}. 
It can be seen that only REM entries being in the close geographical neighborhood can be grouped. The biggest area covered with a single cluster is the forest area, while in the more urbanized regions no clusters could be identified. The DBSCAN provides about $7\%$ REM size reduction. This value is less than, $12\%$ reported previously, due to constraint on grouping together only REM entries being within close geographical distance. Although this gain is moderate, we can expect that gains will increase while measurement density rises. Moreover, as the 2.4 GHz band of consideration presents the most varying interference conditions, it is expected that in another band more homogeneous interference sources will result in higher REM size reduction. 

\section{Evaluation of the REM-based Channel Assignment} \label{sec:simulation}
In order to evaluate the proposed REM-based channel assignment method, a  10-truck platoon is considered, traveling from Poznań to Kórnik and back, as it is depicted in Fig.~\ref{fig:route_map}. To obtain numerical results, a 
 well-established, two slope path loss model is used~\cite{cheng2007}. Through analysis of measurement data authors of the model observed that path loss exponent is rapidly changing after distance between two vehicles $d$ exceeds the critical value $d_{\mathrm{c}}$. 
The received signal strength (RSS) observed by the receiver at distance $d$ from the transmitter is given by:
\begin{equation}
    P(d) = 
    \begin{cases} P(d_{0}) -10\gamma_1 \log_{\mathrm{10}}(\frac{d}{d_0}) + X_{\sigma_1} &\text{  if } d_0 \leq d \leq d_{\mathrm{c}} \\
    P(d_{0}) -10\gamma_1 \log_{\mathrm{10}}(\frac{d_{\mathrm{c}}}{d_0})  &\text{  if } d > d_{\mathrm{c}}\\ \hspace{1cm} -10\gamma_2 \log_{\mathrm{10}}(\frac{d}{d_{\mathrm{c}}}) + X_{\sigma_2}
    \end{cases},
\end{equation}
where $\gamma_1$ and $\gamma_2$ are path loss exponents, $X_{\sigma_1}$, and $X_{\sigma_2}$ are zero-mean Gaussian distributed random variables that model the effect of shadowing, and $P(d_{0})$ is the RSS at the reference distance $d_0$. We assume the $P(d_{0})$ is computed with the use of the Free Space Loss model, for the reference distance of $d_0 =1$~m.

The set of parameters used by the platoon is presented in Tab.~\ref{tab:parameters}. The maximal inter-vehicles distance in a single platoon is set to 200~m. It is an approximate length of a 10-truck platoon. Transmitted power per subcarrier $\mathrm{P_{tx}}$ value is the maximum allowed in 2.4~GHz band~\cite{etsiWlan}. This includes antenna gains. The number of sub-carriers and sub-carrier spacing follows the 802.11p standard~\cite{Abdelgader2014ThePL}. Finally, 3~Mbit/s is set to be the desired capacity $C_{\mathrm{th}}$. It is the lowest supported bitrate in 802.11p, claimed to be used for emergency messages~\cite{campolo2013}. The radio channel parameters are $d_{\mathrm{c}}=100$ m, $\gamma_1$ = 2, $\gamma_2$ = 4, $\sigma_{1} = 5.6$ dB, and $\sigma_2 = 8.4$ dB. These are proper for the considered 2-lane street scenario~\cite{cheng2007}.
\begin{table}[!t]
\renewcommand{\arraystretch}{1.3}
\caption{Platoon configuration and radio parameters 
}
\label{tab:parameters}
\centering
\begin{tabular}{|c|c|}
\hline
Parameter & Value \\
\hline
frequency $f$ & $2.4$ GHz \\ distance $d$ & $200$ m\\ sub-carrier spacing $\mathrm{B}$ & $156.3$ kHz \\
number of used sub-carriers $\mathrm{N_f}$ & 48 \\
transmitted power per sub-carrier $\mathrm{P_{tx}}$ & $20$ dBm $-10\log_{10}(\mathrm{N_f})$  \\
desired capacity $C_{\mathrm{th}}$ & $3$ Mbit/s \\
critical distance $d_{\mathrm{c}}$ & 100 m \\
path loss exponent $\gamma_1$ & 2 \\
path loss exponent $\gamma_2$ & 4 \\
$X_{\sigma_1}$ standard deviation $\sigma_1$ & 5.6 dB\\
$X_{\sigma_2}$ standard deviation $\sigma_2$ & 8.4 dB\\
\hline
\end{tabular}
\end{table}

\begin{figure}[!t]
\centering
\includegraphics[width=3.85in]{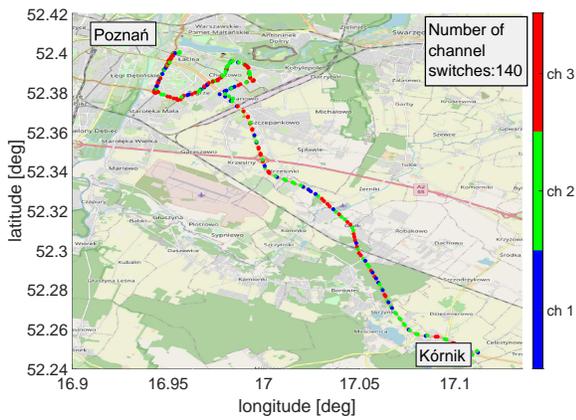}
\caption{Results of channel assignment performed independently in every platoon location using~\eqref{eq:optimal_channel_assignment}.}
\label{fig:ch_assignment_optimal}
\end{figure}
\begin{figure}[!t]
\centering
\includegraphics[width=3.8in]{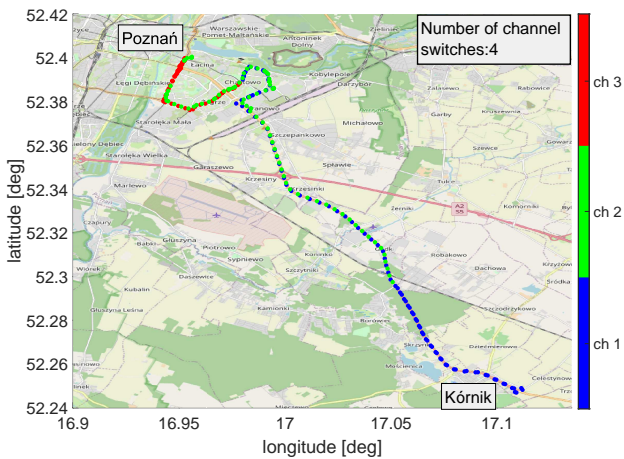}
\caption{Results of channel assignment based on the Dijkstra algorithm.}
\label{fig:ch_assignment_heuristic}
\end{figure}
\begin{figure*}[!t]
\centering
\includegraphics[width=7.6in, height=3in]{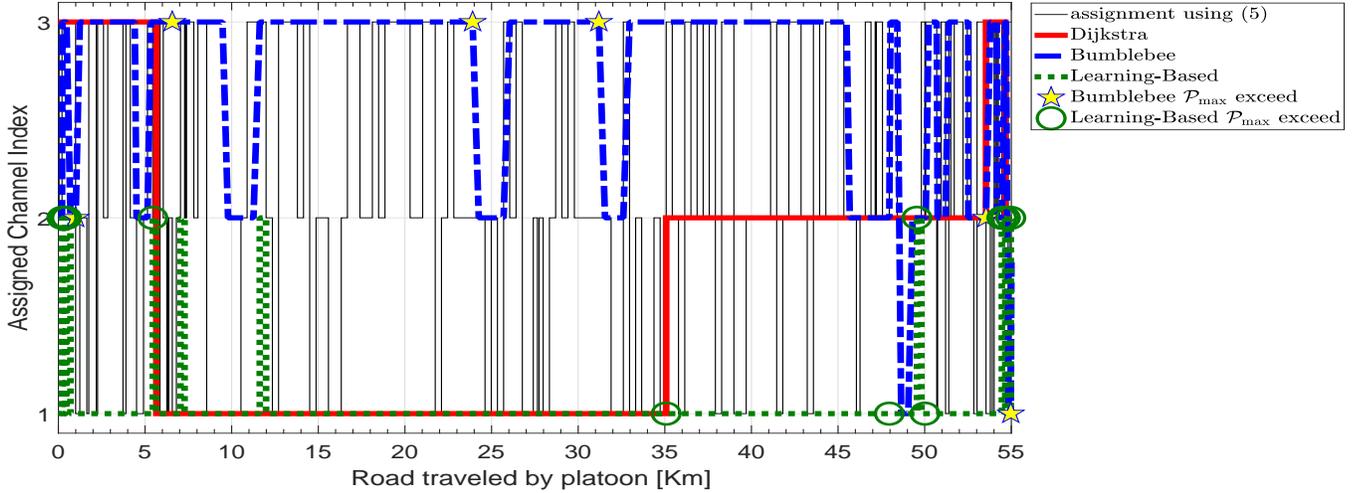}
\caption{Results of the channel assignment along the platoon road.}
\label{fig:ch_assignment_heuristic2}
\end{figure*}

Now the algorithm, where the channel providing the lowest outage probability is chosen for platoon communication in every location according to~\eqref{eq:optimal_channel_assignment} is evaluated. The assigned channels are depicted on the map and as a function of distance from start in Fig.~\ref{fig:ch_assignment_optimal} and Fig.~\ref{fig:ch_assignment_heuristic2}, respectively. It can be easily observed, that this scheme encourages frequent channel switching 
The channel is changed $140$ times along the route. As it was mentioned in Sec.~\ref{subsec:radio_channel_assignment}, the number of channel switches can be reduced, as it is necessary only to provide a sufficient level of outage probability. Following the acceptable bit error rate
in~\cite{campolo2013}, we set maximum allowed outage probability to: $\mathcal{P}_{\mathrm{max}}=10^{-4}$. With the use of this value, the channel assignment along the planned platoon route has been optimized using the Dijkstra algorithm (see Sec.~\ref{subsec:radio_channel_assignment}). The obtained results are depicted in Fig.~\ref{fig:ch_assignment_heuristic}, and Fig.~\ref{fig:ch_assignment_heuristic2}, respectively. With the use of the Dijkstra algorithm, the number of channel switches has been drastically decreased. Instead of $140$ there are only $4$ channel switches along the route, which stands for almost $35$ times reduction. 

Finally, we have compared the proposed solutions against two state-of-the-art algorithms aimed at dynamic channel selection for vehicular communications. First 
algorithm follows bumblebees behaviour~\cite{kuldeep2018}. This will be called a \emph{Bumblebee} algorithm here. The algorithm observes the mean interference power variation in the chosen channel over time. 
If the mean interference power observed in the current location is $15\%$ greater than the one observed in the previous location, the channel is switched to the best one, i.e., of the lowest mean interference power. The $15\%$ cost should prevent frequent channel switching.
The second state-of-the-art algorithm utilizes machine learning techniques in order to obtain statistics of channel occupancy~\cite{chen2011}. We will refer to this algorithm as \emph{Learning-Based}. Similar to the proposed approach, \emph{Learning-Based} algorithm utilizes a location-dependent database. However, instead of direct modeling of channel characteristics, channel occupancy is translated onto the reward. The authors of \cite{chen2011} assumed the arbitrary values of the reward $3$ for the free channel and $-3$ for the busy channel. In every platoon location, channel statistics are an exponential average of the reward. We have trained the \emph{Learning-Based} algorithm with the use of our measurement data, and under the assumption that the channel is busy when $\mathcal{P}_{\mathrm{max}}$ is exceeded.

\begin{figure}[!t]
\centering
\includegraphics[width=3.8in]{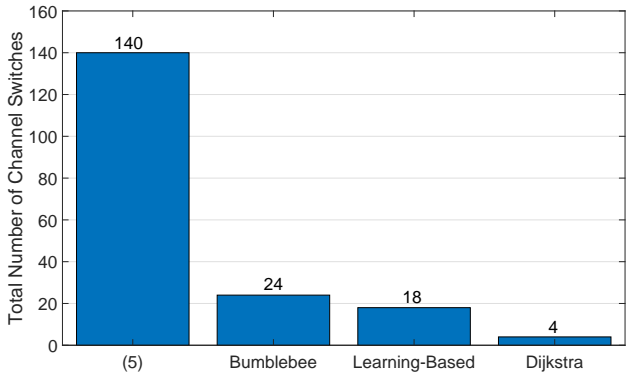}
\caption{Total number of channel switches along the platoon road.} 
\label{fig:no_channel_switches}
\end{figure}

Comparison between channels selected along the platoon road, by the proposed algorithm based on equation~\eqref{eq:optimal_channel_assignment}, proposed application of Dijkstra algorithm, \emph{Bumblebee} algorithm, and \emph{Learning-Based} algorithm is depicted in Fig.~\ref{fig:ch_assignment_heuristic2}. The most important information that can be observed is that both \emph{Bumblebee} algorithm (yellow stars) and \emph{Learning-Based} algorithm (green circles) sometimes chose channels that exceed the allowable level of outage probability $\mathcal{P}_\mathrm{max}$, i.e., $6$, and $12$ times respectively. It is due to modeling inaccuracies. \emph{Bumblebee} utilizes mean interference power to assess the quality of radio channel and \emph{Learning-Based} algorithm stores an exponential average of the reward, not interference distribution itself. Instead, the proposed Dijkstra algorithm utilizes REM, storing information about interference distribution that prevents the outage for selected channels from exceeding $\mathcal{P}_{\mathrm{max}}$. Furthermore, it can be seen that both \emph{Bumblebee}, and \emph{Learning-Based} algorithms do not switch channels as often as in the case of channel assignment based on~\eqref{eq:optimal_channel_assignment}. outperforms them. This relation is clearly visible in Fig.~\ref{fig:no_channel_switches}. In the case of \emph{Bumblebee}, and \emph{Learning-Based} algorithm channel switching occurs 24, and 18 times, respectively. However, when the Dijkstra algorithm is used the platoon switches channel only 4 times. It stands for the 6 times reduction in relation to the \emph{Bumblebee} algorithm and $4.5$ times reduction in relation to the \emph{Learning-Based} algorithm. It is because the Dijkstra algorithm optimizes the channel switching globally along the whole route using knowledge from REM. The \emph{Bumblebee} algorithm utilizes only locally available sensing samples. Although \emph{Learning-Based} algorithm also utilizes a database with location-dependent data, the channel selection algorithm is designed so as it does not take into account the dependencies between consecutive locations like our solution does. 

\begin{figure}[!t]
\centering
\includegraphics[width=3.8in]{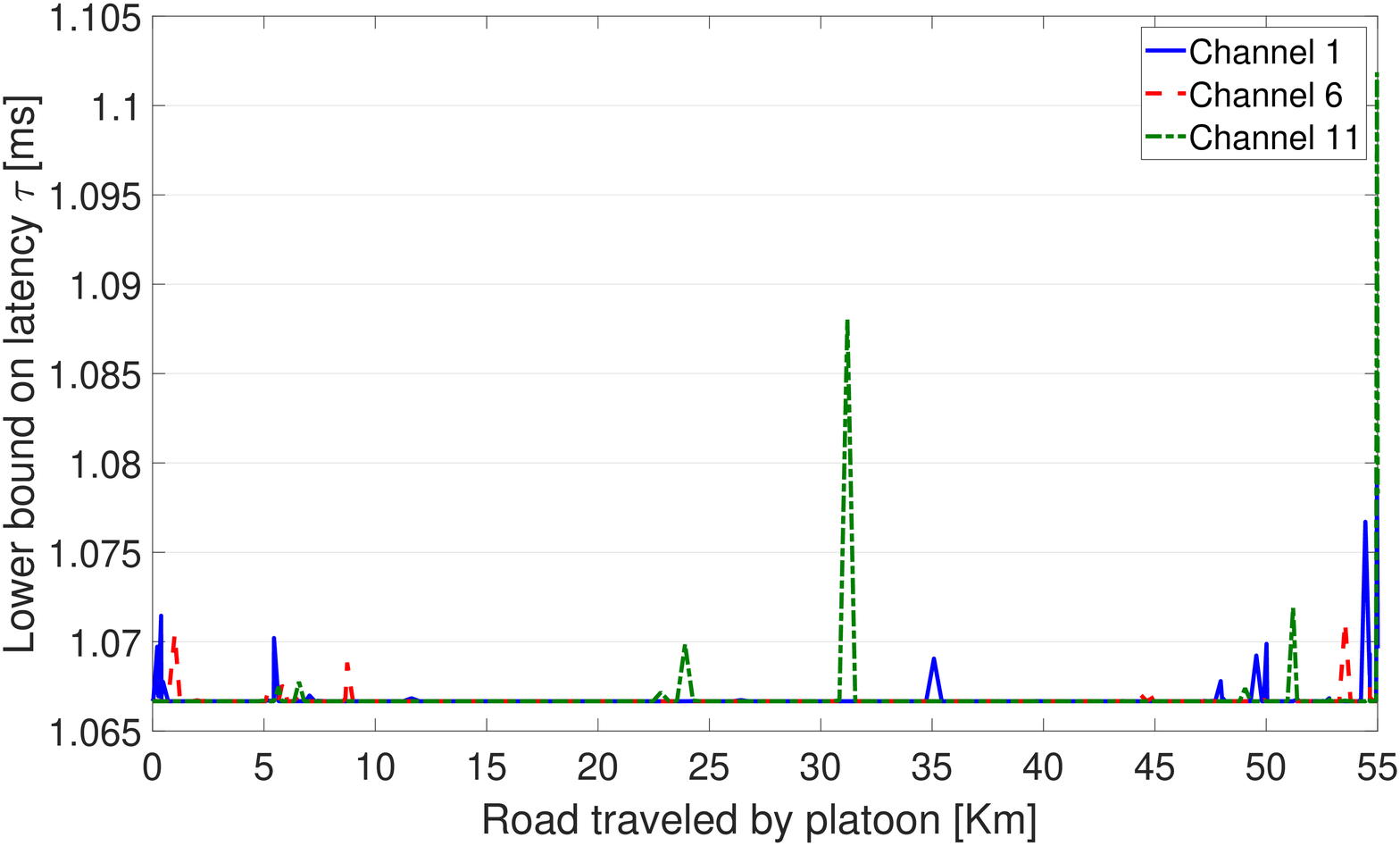}
\caption{Lower bound of latency along the platoon road for packet size $D=400$~bytes.}
\label{fig:latency}
\end{figure}

\begin{figure}[!t]
\centering
\includegraphics[width=3.8in]{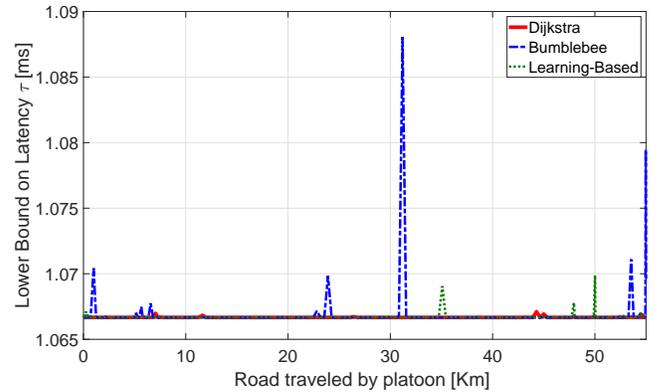}
\caption{Lower bound of latency along the platoon road for different channel selection algorithms, and packet size of $D=400$~bytes.}
\label{fig:latency2}
\end{figure}

Finally, Fig.~\ref{fig:latency} shows a lower bound of latency computed using~\eqref{eq:latency} along the platoon road, for the packet size $D = 400$~bytes~\cite{REDDYG2018720}. It can be observed that in most of the locations latency is on the level of $1.067$~ms. This is because in those locations outage probability is close to 0, and a fixed throughput equal to channel capacity $C_{\mathrm{th}}$ is assumed in \eqref{eq:latency}. Although in the locations where outage probability is the highest the lower bound latency increases by up to 35~$\mu$s (by $3$\%), in practice delays are expected to be much longer, e.g., due to the specific medium access algorithm. Still, this $3$\% increase can have a non-negligible, negative impact on the platoon control system.
In Fig.~\ref{fig:latency2} there is a comparison of the lower bound of latency between the proposed Dijkstra algorithm, and two state-of-the-art solutions: \emph{Bumblebee} algorithm and \emph{Learning-Based} algorithm. We can see that the proposed algorithm is characterized by the lowest lower bound of latency oscillating around the level of $1.0667$~ms, and not exceeding the level of $1.0671$~ms. The highest peaks could be observed for the Bumblebee algorithm, it is caused by the modeling inaccuracies, i.e., channel selection based on the mean interference power. The performance of \emph{Learning-Based} algorithm is in-between these two algorithm mentioned above. Still some latency peaks can be observed. This is because \emph{Learning-Based} algorithm utilizes better model of channel characteristic than \emph{Bumblebee} algorithm, but not as good as the proposed GMM model.           

\section{Conclusions} \label{sec:conclusions}
In this paper, we have presented the design of REM with aim of selecting a proper secondary spectrum channel for intra-platoon communications. This included a method for assessment of available secondary spectrum channels quality, application of Dijkstra algorithm for frequency selection in order to reduce the number of channel switches along the platoon route, and utilization of DBSCAN for the purpose of REM size reduction.

The observed non-trivial distributions of interference can be effectively modeled with GMM of 5 components as proved using AIC. This GMM interference model can be effectively used to construct REM. The size of REM can be reduced by grouping REM entries of similar interference distributions, and being in close geographical distance. It can be done using the proposed modification to the DBSCAN algorithm, and provide the reduction of REM size of about $7\%$. Finally, the assignment of the secondary spectrum channel independently in every platoon location results in a high number of channel switches. The proposed, Dijkstra algorithm-based method reduces the number of channel switches about $35$ times.

In the future, additional measurement campaigns are necessary, e.g., to observe how GMM parameters are changing over the daytime. Having more data captured over different daytime, algorithms for recursive GMM updates can be evaluated. Also, interference distribution should be evaluated under other frequency bands potentially promising for opportunistic intra-platoon communication e.g., TVWS, WiFi\texttrademark{} 5~GHz band. Finally, an advanced network simulator can be developed to study the proposed REM algorithms under realistic conditions, where also interference from other platoons will be taken into the account.

\ifCLASSOPTIONcaptionsoff
  \newpage
\fi



\bibliographystyle{IEEEtran}
\bibliography{bibliography.bib}
%

%
\begin{IEEEbiographynophoto}{Marcin Hoffmann} received  his M.Sc. degree (with honors)  in Electronics and Telecommunication from Poznań University of Technology in 2019. Since October 2019 He’s a Ph.D. student in the Institute of Radiocommunications at Poznań University of Technology.  He is gaining scientific experience by being involved in both national, and international research projects.  His main fields of interests is utilization of machine learning and location-dependent information for the purpose of network management.
\end{IEEEbiographynophoto}
\begin{IEEEbiographynophoto}{Pawel Kryszkiewicz} received the M.Sc. and Ph.D. degrees (Hons.) in telecommunications from the Poznan University of Technology (PUT), Poland, in 2010 and 2015, respectively. He is
currently an Assistant Professor with the Institute of Radiocommunciations, PUT. He was involved in a number of national and international research projects. His main fields of interest are problems concerning the physical layer of the
Dynamic Spectrum Access system, multicarrier signal design for green communications, and problems related to practical implementation of Massive MIMO systems.
\end{IEEEbiographynophoto}
\begin{IEEEbiographynophoto}{Adrian Kliks}
is an associate professor at Poznan University of Technology’s Institute of Radiocommunications, Poland. His research interests include new waveforms for wireless systems applying either nonorthogonal or noncontiguous multicarrier schemes, cognitive radio, advanced spectrum management, deployment and resource management in small cells, and network virtualization.
\end{IEEEbiographynophoto}


\vfill


\end{document}